\def\rfr#1{eq. (\ref{#1})}

\def\cf#1#2{\dot\Omega^{\rm #2}_{.#1}}

\def\derp#1#2{\rp{\partial{#1}}{\partial{#2}}}
\def\bar{\begin{eqnarray}}
\def\ear{\end{eqnarray}}
\def\bb{\bibitem}
\def\eqi{\begin{equation}}
\def\eqf{\end{equation}}
\def\eqia{\begin{eqnarray}}
\def\eqfa{\end{eqnarray}}
\def\rp#1#2{{#1\over#2}}

\def\lb#1{\label{#1}}

\def\oc2{$\mathcal{O}(c^{-2})$}

\documentclass[11pt]{article}
\usepackage{amsmath,amsthm,amscd,amssymb}
\usepackage{latexsym}
\usepackage{graphicx,epsfig}

\begin{document}

\noindent{\bf \LARGE{Mars and frame-dragging: study for a dedicated mission }}
\\
\\
\\
{Lorenzo Iorio}\\
{\it INFN-Sezione di Pisa}\\
{\it Viale Unit$\grave{\it a}$ di Italia 68, 70125\\Bari (BA), Italy
\\tel. 0039 328 6128815
\\e-mail: lorenzo.iorio@libero.it}

\begin{abstract}
In this paper we preliminarily explore the possibility of designing a dedicated
satellite-based mission to measure the general relativistic gravitomagnetic Lense-Thirring effect
in the gravitational field of Mars. The focus is on the systematic error induced by the multipolar expansion of the areopotential and on possible strategies to reduce it.  It turns out that the major sources of bias are the Mars'equatorial radius $R$ and the even zonal harmonics $J_{\ell}, \ell=2,4,6...$ of the areopotential. An optimal solution, in principle,  consists of using two probes at high-altitudes ($a\approx 9500-9600$ km) and different inclinations (one probe should fly in a nearly polar orbit), and suitably combining their nodes in order to entirely  cancel out the bias due to $\delta R$. The remaining uncancelled mismodelled terms due to $\delta J_{\ell},\ell=2,4,6,...$ would induce a bias $\lesssim 1\%$, according to the present-day MGS95J gravity model, over a wide range of admissible values of the inclinations. The Lense-Thirring out-of-plane shifts of the two probes would amount to about $10$ cm yr$^{-1}$.
 \end{abstract}



\section{Introduction}
Recent claims concerning a detection of the general relativistic gravitomagnetic Lense-Thirring effect \cite{LT}  in the gravitational field of Mars with the Mars Global Surveyor probe \cite{MGS1,MGS2,krogh,reply} raised interest concerning such a new Solar System scenario for testing Einsteinian gravity. In view of what happened to the Gravity Probe B mission \cite{Eve,GPB}, which might finally reach an accuracy\footnote{See on the WEB http://einstein.stanford.edu/} far from the originally expected $1\%$ in measuring the Schiff precessions \cite{Schi} of the spins of four gyroscopes carried onboard, and of the lingering controversy about the realistic accuracy reached with the terrestrial LAGEOS-LAGEOS II test of the Lense-Thirring effect \cite{IorPSS} (and reachable with the recently approved LAGEOS-like LARES satellite, to be launched at the end of\footnote{See on the WEB http://www.esa.int/esapub/bulletin/bulletin135/bul135f$\_$bianchi.pdf} 2009 by the Italian Space Agency \cite{Ior08}) we feel that attempts to scour new and unexplored routes may be of some value.
Indeed, as many laboratories and methods as possible should be used to extensively test a fundamental interaction like gravity and its predictions, especially when their tests are so few and their outcomes uncertain.

In this paper we wish to fix a first stick by studying  in some details the possibility of using a dedicated mission to Mars with one or more probes to measure at a reasonable level of accuracy (a few percent) the elusive Lense-Thirring effect on the longitude of the ascending node $\Omega$ of the spacecraft's orbital plane. We will concentrate here on one of the major source of  systematic error, i.e. the multipolar expansion of the martian gravitational potential which induces on the node a huge noise with the same temporal signature of the relativistic signal of interest (a linear rate) in order to see what are the critical issues in view of the present-day knowledge of the Martian space environment.  We will not discuss here the perturbations of non-gravitational origin which depend on the shape, the instrumentation and the orbital maneuvers  of such  probes. They would be strongly  related to possible other tasks, more consistent with planetology, which could be fruitfully assigned to such a mission in order to enhance the possibility that it may become something more than a mere, although-hopefully-interesting, speculation; in this respect the medium-long term ambitious programs of NASA to Mars may turn out to be useful also for the purpose discussed here.
%
%\section{The systematic error of gravitational origin}
\section{The use of one nearly polar spacecraft}
The Lense-Thirring effect consists of a small secular precession of the node\footnote{Also the pericentre $\omega$ of a satellite secularly precesses under the action of the gravitomagnetic force, but we will not consider here such an effect.} $\Omega$ of  the orbit of a satellite moving around a central slowly rotating mass
\eqi\dot\Omega_{\rm LT}=\rp{2GS}{c^2 a^3(1-e^2)^{3/2}},\eqf
where $G$ is the Newtonian constant of gravitation, $S$ is the spin angular momentum of the central body, $a$ and $e$ are the semimajor axis and the eccentricity, respectively, of the satellite orbit.  According to the latest determinations of the global properties of Mars \cite{Kon06}, $S = (1.92\pm 0.01)\times 10^{32}$ kg m$^2$ s$^{-1}$ for the red planet.

The oblateness of the central body, of mass $M$ and equatorial radius $R$, makes the node to secularly precess as well according to \cite{Kau}
\eqi\dot\Omega^{\rm (obl)}=\sum_{\ell=2}\dot\Omega_{.\ell}J_{\ell},\eqf
 where the coefficients $\dot\Omega_{.\ell}$, explicitly computed in, e.g., \cite{book} up to degree $\ell=20$,   depend on the orbital semimajor axis, the eccentricity and the inclination $i$ to the planet's equator;
 for example, for $\ell=2$ we have
 \eqi\dot\Omega_{.2}=-\rp{3}{2}n\left(\rp{R}{a}\right)^2\rp{\cos i}{(1-e^2)^2},\eqf
 where $n=\sqrt{GM/a^3}$ is the probe's Keplerian mean motion.
 Although the multipolar expansion of the classical part of the gravitational potential is usually included in the dynamical force models used to process probes' data, such a  modelling is always imperfect; thus,  an aliasing trend due to the mismodelling in the even zonals  $J_{\ell}$ and the other parameters like $R$ and $GM$ entering the classical node precessions may affect the recovery of the relativistic precession of interest.

\subsection{The even zonal harmonics}
 Let us, now, focus on the systematic error $\delta\mu$ induced by the uncertainty in the even zonals $\delta J_{\ell}$ by assuming for them the calibrated covariance sigmas of the latest Mars gravity model MGS95J \cite{Kon06}.  We will evaluate it as
 \eqi \delta\mu_{J_{\ell}}\leq |\dot\Omega_{.\ell}|\delta J_{\ell},\ \ell=2,4,...\eqf
 Because  of the presence of $\cos i$ in all the coefficients $\dot\Omega_{.\ell}$ of the oblateness-induced node precessions, we will concentrate on nearly polar ($i\approx 90$ deg) orbital configurations to minimize such a corrupting effect.

\begin{figure}
   \includegraphics[width=\columnwidth]{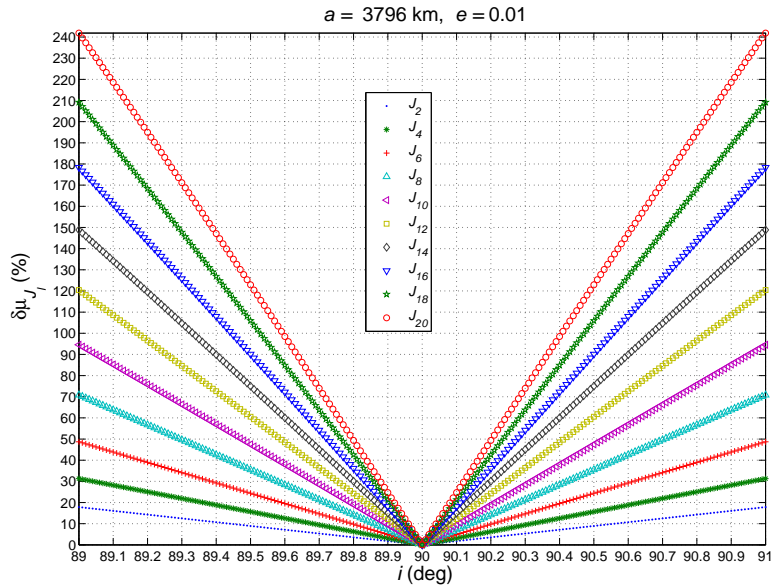}
   \caption{Systematic percent errors $\delta\mu_{J_{\ell}}$ per degree $\ell$ due to the mismodelling in the even zonal harmonics $\delta J_{\ell},\ell=2,4,6...$ for $a=3796$ km, $e=0.01$, $89$ deg $\leq i\leq 91$ deg according to the calibrated covariance sigmas of the MGS95J global gravity solution up to $\ell=20$. The Lense-Thirring effect amounts to 34 mas yr$^{-1}$ corresponding to a shift in the out-of-plane direction of 0.62 m yr$^{-1}$.}
   \label{figura1}
   \end{figure}

 It turns out that  altitudes of a few hundreds of km typical of the majority of the currently ongoing martian missions  are definitely not suited for our scope, as shown by Figure \ref{figura1}.
 Note that inserting a probe into an areocentric orbit is not an easy task, so that we decided to allow for a departure of up to 1 deg from the ideal polar orbital configuration to account for unavoidable orbital injection errors. Also typical mission requirements pull the inclination some degrees apart from 90 deg: for Mars Global Surveyor (MGS) $i=92.86$ deg. The systematic bias per degree increases for higher degrees $\ell=2,4,6,...$ and it turns out that only a very narrow range for $i$, i.e. $\Delta i\approx 10^{-3}$ deg, unlikely to obtain, might push the systematic errors below the $1\%$ level.

\begin{figure}
   \includegraphics[width=\columnwidth]{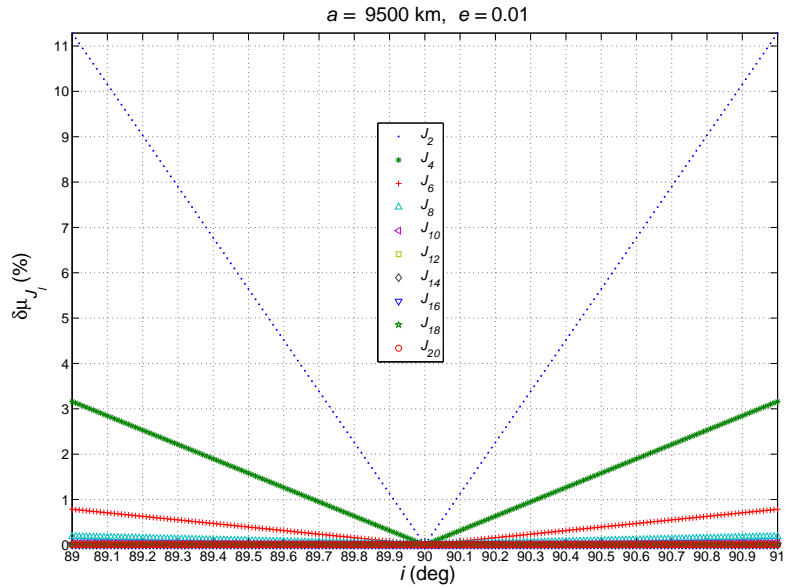}
   \caption{Systematic percent errors $\delta\mu_{J_{\ell}}$ per degree $\ell$ due to the mismodelling in the even zonal harmonics $\delta J_{\ell},\ell=2,4,6...$ for $a=9500$ km, $e=0.01$, $89$ deg $\leq i\leq 91$ deg according to the calibrated covariance sigmas of the MGS95J gravity model up to $\ell=20$. The Lense-Thirring effect amounts to 2 mas yr$^{-1}$ corresponding to a shift in the out-of-plane direction of 0.10 m yr$^{-1}$.}
   \label{figura2}
   \end{figure}

 By keeping a near polar geometry, much larger values of the semimajor axis, comparable with that of Phobos ($a=9380$ km), one of the two natural satellites of Mars, yield reasonable results, as shown by Figure \ref{figura2}.
 Indeed, now the noise decreases with high degree terms and the most serious contribution is due to the first even zonal $\ell=2$ yielding a bias up to about 11$\%$ for $\Delta i=1$ deg; $J_4$ affects the the Lense-Thirring precessions at most at $3\%$ level for $\Delta i=1$ deg. Note also that, now, a much larger departure from 90 deg ($\Delta i\approx 0.1$ deg) would allow a further reduction of the noise ($\lesssim 1\%$).   It must be noted that for such a high-altitude orbital configuration the gravitomagnetic shift in the out-of-plane direction would amount to 0.10 m yr$^{-1}$; it is certainly a small figure, but, perhaps, not too small if one considers that the average shift in the out-of-plane direction of the much lower Mars Global Surveyor is 1.6 m after about 5 yr. It poses undoubtedly challenging requirements in terms of sensitivity and overall orbit determination accuracy, but it seems not unrealistic to hope that future improvements may allow to detect such a small displacement.

 Finally, let us remark that, in principle, also the temporal variations of $J_2$ should be accounted for; however, since at present no secular trends have been detected, such changes, mainly seasonal, annual and semi-annual \cite{Kon06} would not seriously impact our measurement.
\subsection{The impact of $R$ and $GM$}
\begin{figure}
   \includegraphics[width=\columnwidth]{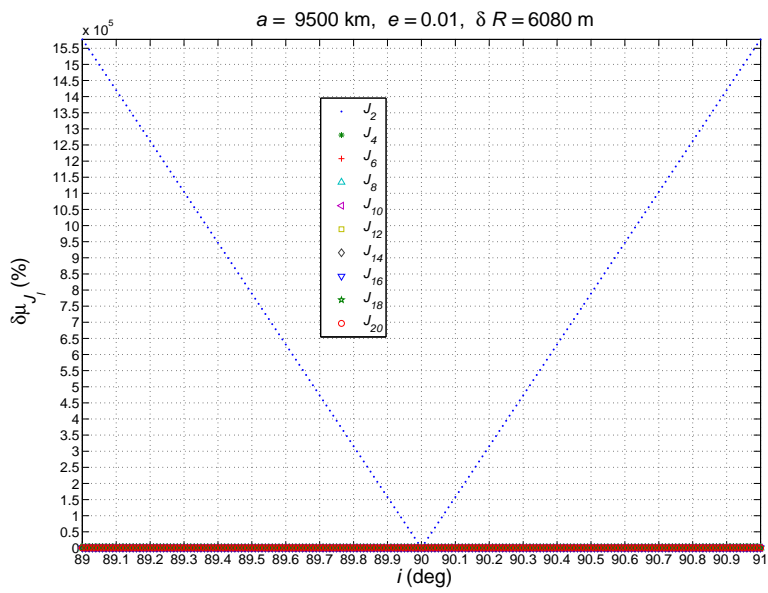}
   \caption{Systematic percent errors $\delta\mu_{J_{\ell}}$ per degree $\ell$ due to the uncertainty in Mars' radius $R$, assumed to be $\delta R = 6080$ m, for $a=9500$ km, $e=0.01$, $89$ deg $\leq i\leq 91$ deg. The errors for $\ell=4,6$ are as large as $6000\%$ and $500\%$, respectively.}
   \label{figura3}
   \end{figure}

\begin{figure} 
 \includegraphics[width=\columnwidth]{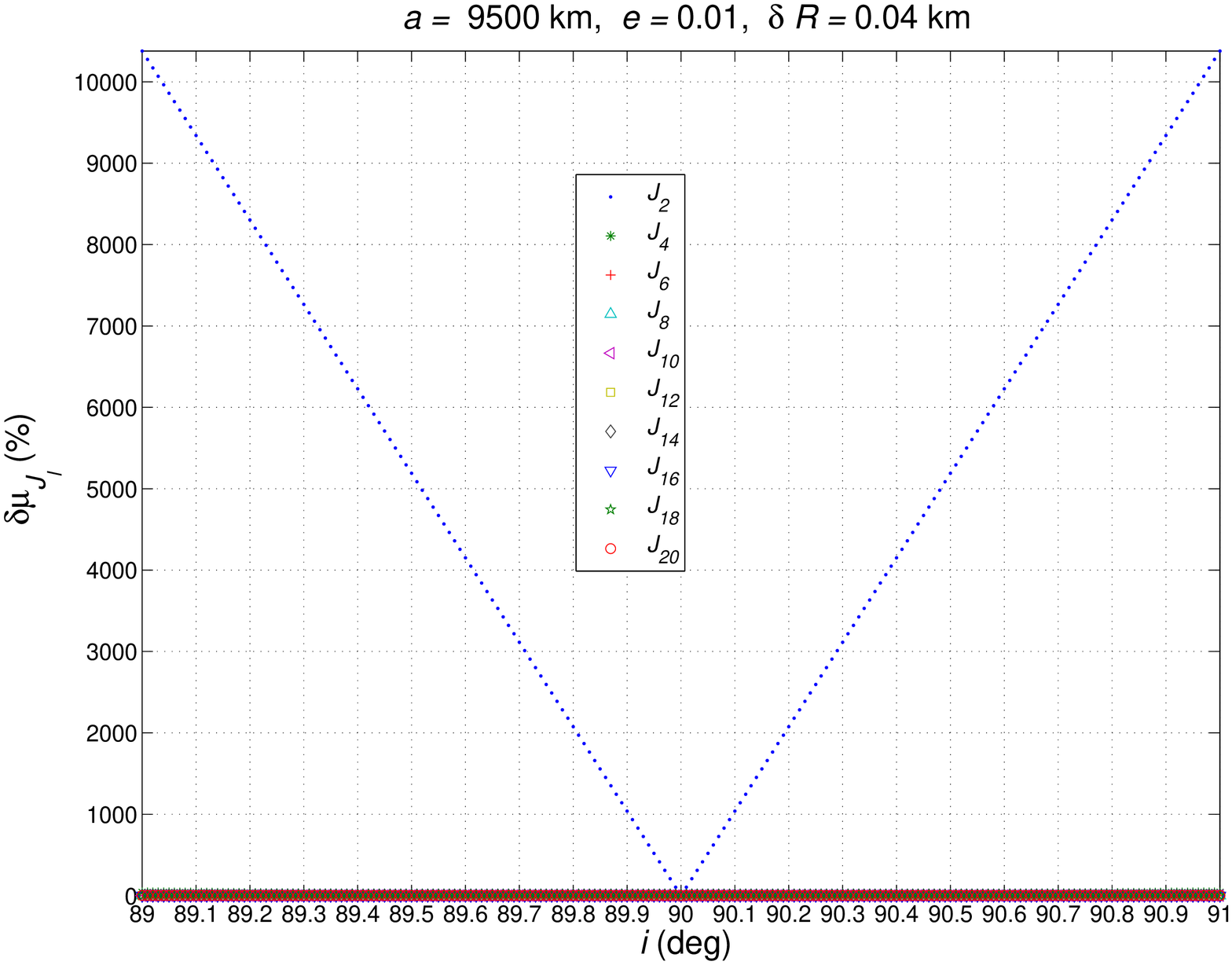}
   \caption{Systematic percent errors $\delta\mu_{J_{\ell}}$ per degree $\ell$ due to the uncertainty in Mars' radius $R$, assumed to be $\delta R = 0.04$ km, for $a=9500$ km, $e=0.01$, $89$ deg $\leq i\leq 91$ deg. The errors for $\ell=4,6$ are as large as $40\%$ and $5\%$, respectively.}
   \label{figura4}
   \end{figure}

Another martian parameter which must be taken into account is the equatorial radius $R$ along with its uncertainty. 
Since
\eqi\dot\Omega_{.\ell}\propto R^{\ell},\eqf the systematic error per degree induced by $\delta R$
can be written as
\eqi\delta\mu_{J_{\ell}}\leq \ell\left(\rp{\delta R}{R}\right)\left|\dot\Omega_{.\ell}J_{\ell}\right|,\ \ell=2,4,6...\eqf
If we assume conservatively for it $\delta R = 6080$ m, i.e. the difference between the reference value of MGS95J \cite{Kon06} and  the one in \cite{Yod95}, the result is depicted in Figure \ref{figura3} for $a=9500$ km.
If, instead, one takes $\delta R=0.04$ km the errors per degree are as in Figure \ref{figura4} for $a=9500$ km.
As can be noted,  $R$ is the most serious source of systematic error, even for high altitudes.  Moreover, it must also be considered that it would be likely unrealistic to expect improvements in the determination of $R$ by future areocentric missions able to push $\delta R$ below terrestrial values, i.e. $1-0.1$ m.  Thus,
$R$ will likely remain an insurmountable obstacle if only one probe is to be used.

Concerning $GM$, it turns out that the impact of its mismodelling   is of no concern being $\lesssim 1\%$.
 %
 %
 %
 %\begin{figure}[htbp]
 %  \includegraphics[width=17cm,height=13cm]{MARSGM.jpg}
 %  \caption{Systematic percent errors $\delta\mu$ per degree $\ell$ due to the uncertainty in Mars' $GM$, assumed to be $\delta GM = 2.8\times 10^5$ %m$^3$ s$^{-2}$ \cite{Kon06}, for $a=9500$ km, $e=0.01$, $89$ deg $\leq i\leq 91$ deg.}
 %  \label{figura5}
 %  \end{figure}
 %
 %
 %
\subsection{Summary}
In summary, in this Section we investigated a nearly polar, high-altitude Phobos-like orbit and noted that the Lense-Thirring effect amounts to a shift of 0.10 m yr$^{-1}$ in the out-of-plane direction. Concerning the systematic errors, the
most crucial source of aliasing is the radius $R$ ($\delta\mu\approx 10000 \%$ for $\delta R = 0.04$ km) and, to a lesser extent, the first even zonal harmonic $J_{2}$ of the areopotential ($\delta\mu\leq 11\%$). To  reduce them the probe should be inserted into an orbit with an inclination close to 90 deg within $10^{-3} $ deg or less. In conjunction with such a very tight constraint, it must also be hoped that future missions to Mars will improve our knowledge of the fundamental parameters of the red planet to a sufficient extent to allow for larger departures from the ideal polar geometry: after all, the gravity solution MGS95J represents an improvement of one order of magnitude with respect to the previous MGS75D model \cite{Yuan01}. However, this may be valid for the spherical harmonic coefficients of the areopotential, not for the radius of Mars for which an uncertainty of the order of the meter would not be enough. We must conclude that the use of only one probe for measuring its node Lense-Thirring precession is unfeasible.

\section{Including Phobos and Deimos}
A possible solution to reduce the systematic errors would be to consider a linear combination of the nodes of the proposed probe and of the two natural satellites of Mars, i.e. Phobos ($a = 9380$ km, $i = 1.075$ deg, $e = 0.0151$) and Deimos ($a = 23460$ km, $i = 1.793$ deg, $e = 0.0002$), suitably designed to cancel out the impact of just $\delta J_2$ and $\delta R/R$, according to an approach put forth for the first time in the context of the terrestrial LAGEOS-LAGEOS II test \cite{Ciu96}.
A possible combination  is
\eqi \delta\dot\Omega^{\rm Deimos} + c_1\delta\dot\Omega^{\rm Probe}+c_2\delta\dot\Omega^{\rm Phobos},\lb{combi}\eqf
with $\delta\dot\Omega$ denoting some sort of Observed-minus-Calculated ($O-C$) quantity  accounting for any unmodelled/mismodelled features of motion and extracted from the data processing with full dynamical force models. By purposely leaving the gravitomagnetic force unmodelled, it can be written down as
\eqi\delta\dot\Omega = \mu X_{\rm LT} + \dot\Omega_{.2}\delta J_{2} + \dot\Omega_{.R}\left(\rp{\delta R}{R}\right) + \Delta:\eqf
here $\mu=1$ in general relativity,   the coefficient $\dot\Omega_{.R}$ is defined as
\eqi \dot\Omega_{.R} = R\derp{\dot\Omega^{\rm (obl)}}{R}=\sum_{\ell=2}\ell\dot\Omega_{.\ell}J_{\ell},\eqf
and $\Delta$ represents all other remaining mismodelled/unmodelled effects acting on the node (e.g. $\delta J_4, \delta J_6,...; \delta GM$). By solving for $\mu$ one gets
  \begin{equation}
\left\{
\begin{array}{lll}
c_1 = \rp{\cf 2{\rm Phobos}\dot\Omega_{.R}^{\rm Deimos}-\cf 2{\rm Deimos}\dot\Omega_{.R}^{\rm Phobos}}{\cf 2{\rm Probe}\dot\Omega_{.R}^{\rm Phobos}-\cf 2{Phobos}\dot\Omega_{.R}^{\rm Probe}},\\\\
c_2 =  \rp{\cf 2{\rm Deimos}\dot\Omega_{.R}^{\rm Probe}-\cf 2{\rm Probe}\dot\Omega_{.R}^{\rm Deimos}}{\cf 2{\rm Probe}\dot\Omega_{.R}^{\rm Phobos}-\cf 2{\rm Phobos}\dot\Omega_{.R}^{\rm Probe}}.
\end{array}\lb{cofi}
\right.
\end{equation}
Note that $c_1$, contrary to $c_2$, is not defined for $i=90$ deg because $\dot\Omega_{.\ell}^{\rm Probe}$, which are proportional to $\cos i$, vanish for all $\ell$ at $i=90$ deg.
The result, in principle, would be satisfactory. Indeed,
 %
 %
 %
 %\begin{figure}[htbp]
 %  \includegraphics[width=17cm,height=13cm]{MARSmiracolo.jpg}
 %  \caption{Systematic percent errors $\delta\mu$ per degree $\ell$ due to the mismodelling in $J_4, J_6, J_8...$ (MGS95J model), for a combination %with the nodes of Phobos and Deimos designed to cancel out $\delta J_2$ and $\delta R/R$. The orbital elements of the probe are $a=9500$ km, %$e=0.01$, $89$ deg $\leq i\leq 89.9$ deg. For $i>90$ deg the errors are larger.}
 %  \label{figura6}
 %  \end{figure}
 %
 %
 %
now, the systematic bias of $J_4, J_6, J_8...$ on the combination of \rfr{combi} turns out to be $\delta\mu\lesssim 6\%$ in $89$ deg $\leq i\leq 89.9$ deg; the major contribution is due to $J_4$. It is an acceptable result, especially, in view of the fact that the admissible range for the inclination amounts, now,  to about 1 deg and that \rfr{combi} is, by construction, immune to the uncertainty in the martian radius; indeed, although it is likely that many physical properties and parameters like, e.g., the even zonals, of the red planets will be determined with increasing accuracy by the
many ongoing and planned missions, it is unlikely that the radius will be known to a sufficient accuracy to change at an acceptable level the bias induced by it (see Figure \ref{figura3} and Figure \ref{figura4}).
%It turns out that inclinations larger than 90 deg would increase $\delta\mu$ with respect to Figure \ref{figura6}.

The combination of \rfr{combi} is also able to remove the large part of the effect of the mismodelling in $GM$
\eqi\delta\mu_{GM}\leq\rp{1}{2}\left(\rp{\delta GM}{GM}\right)\sum_{\ell=2}^{20}\left|\dot\Omega_{.\ell}J_{\ell}\right|.\lb{ergm}\eqf
Indeed, the main contribution to \rfr{ergm}  is due to $J_2$, which is canceled out by \rfr{combi} with the coefficients of \rfr{cofi}.  The other terms in \rfr{ergm}, not canceled by the combination of \rfr{combi}, yield negligible errors, well below $1\%$.

However, it must be stressed that the node of Phobos   undergoes secular precessions due to  other perturbations of gravitational origin \cite{Lai07} (non-sphericity of Phobos itself, Mars tidal bulge and nutation) which have not been considered here and that would affect the combination of \rfr{combi}. It turns out that the most relevant one is that due to the spherical harmonic coefficients $c_{20}$ and $c_{22}$ of Phobos itself \cite{Bor90} whose induced secular precession amounts nominally to 200 km over 3 yr \cite{Lai07}. The nutation perturbation nominally amounts to 0.3 km over 3 yr \cite{Lai07}. The tidally induced precession, parameterized in terms of the martian Love number $k_2$ would amount nominally 0.06 km after 3 yr \cite{Lai07}, but being $k_2=0.152\pm 0.009$   \cite{Kon06}, its modelling would left a $\approx 1$ m yr$^{-1}$ mismodelled trend.
The Phobos gravitomagnetic shift is 0.1 m yr$^{-1}$.

In terms of the detectability of the Lense-Thirring signal with the combination of \rfr{combi}, undoubtedly, a major drawback of the strategy of including Phobos and Deimos is the poor accuracy with which their orbits can be reconstructed with respect to their Lense-Thirring signal. Indeed, while their gravitomagnetic shifts are of the order of $1-10$ cm yr$^{-1}$, the latest NASA martian ephemerides\footnote{See on the WEB http://ssd.jpl.nasa.gov/?sat$\_$ephem.} for them yield an accuracy of $1-10$ km in the radial, transverse and out-of-plane orbital components \cite{MAR}. Thus, according to the present-day level of accuracy    including the natural satellites of Mars in the combination of \rfr{combi} would introduce a noise which would overwhelm the relativistic trend of interest.
 \section{Two dedicated probes}
In principle, the linear combination approach could be followed by using the nodes of other two spacecrafts\footnote{For a polar geometry the Lense-Thirring precession of the pericentre $\omega$ of a spacecraft vanishes being proportional to $\cos i$.}, although sending to Mars three new probes would increase the costs and the difficulties   of such a demanding mission.

\subsection{The supplementary orbit configuration}\lb{supl}
It is interesting to consider a scenario involving only two  probes, named P1 and P2, in a nearly polar counter-orbiting configuration, proposed for the first time by Van Patten and Everitt \cite{vanpa} in the framework of the attempts to design a suitable terrestrial mission, and later generalized by Ciufolini \cite{Ciu86} for other inclinations.
By assuming for, say, P1 $a_1=9500$ km, $i_1=89$ deg, $e_1=0.01$, it turns out that the sum of their nodes would be an observable relatively insensitive to departures from the ideal configuration for P2, i.e.  $a_2=9500$ km, $i_2=91$ deg, $e_2=0.01$, at least in regard to the mismodelling in the even zonals. Indeed, it turns out that  for   $a_2=9600$ km, $90.05$ deg $\leq i_2\leq 92$ deg, $e_2=0.03$ the bias due to $J_2$ is up to $7\%$,
while the other even zonals would have an impact of the order of $1\%$ or less.
%as shown by Figure \ref{figura8}.
%
 %
 %
 %\begin{figure}[htbp]
 %  \includegraphics[width=17cm,height=13cm]{MARSsumnod1.jpg}
 %  \caption{Systematic errors per degree due to the mismodelling $\delta J_\ell$ in the even zonals in the sum of the nodes of probe P1 ($a_1=9500$ %km, $i_1=89$ deg, $e_1=0.01$) and probe P2 for $a_2=9600$ km, $90.05$ deg $\leq i_2\leq 92$ deg, $e_2=0.03$.}
 %  \label{figura8}
 %  \end{figure}
 %
 %
 %
%As can be noted, departures of $1-2$ deg and 100 km from the ideal configuration would still allow for good results.
Unfortunately, the uncertainty in the martian radius makes such constraints much more stringent. Indeed, $\delta\mu_R\approx 10^5\%$ for $\delta R=6080$ m and $\delta\mu_R\approx 4000\%$  for $\delta R=0.04$ km. Also in this case, the improvements in determining $R$ that could realistically be expected from future missions to the red planets may be not sufficient.
 %
 %
 %
 %\begin{figure}[htbp]
 %  \includegraphics[width=17cm,height=13cm]{MARSsumnod2.jpg}
 %  \caption{Systematic errors per degree due to the mismodelling $\delta R=6080$ m in the Mars' radius $R$ in the sum of the nodes of probe P1 %($a_1=9500$ km, $i_1=89$ deg, $e_1=0.01$) and probe P2 for $a_2=9600$ km, $90.5$ deg $\leq i_2\leq 91.5$ deg, $e_2=0.03$. The bias for $\ell=4,6,8$ %are up to 1000$\%$, $200\%$ and $6\%$, respectively.}
 %  \label{figura9}
 %  \end{figure}
 %
 %
 %
 %
 %
 %
 %\begin{figure}[htbp]
 %  \includegraphics[width=17cm,height=13cm]{MARSsumnod3.jpg}
 %  \caption{Systematic errors per degree due to the mismodelling $\delta R=0.04$ km in the Mars' radius $R$ in the sum of the nodes of probe P1 %($a_1=9500$ km, $i_1=89$ deg, $e_1=0.01$) and probe P2 for $a_2=9600$ km, $90.5$ deg $\leq i_2\leq 91.5$ deg, $e_2=0.03$. The bias for $\ell=4,6$ are %up to $16\%$ and $1\%$, respectively.}
 %  \label{figura10}
 %  \end{figure}
 %
 %
 %
\subsection{The linear combination approach}
\begin{figure}
   \includegraphics[width=\columnwidth]{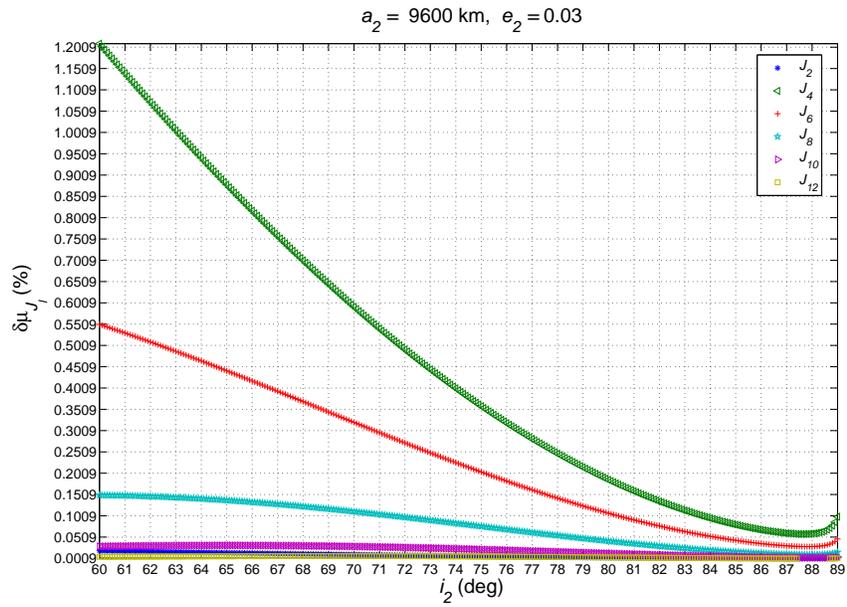}
   \caption{Systematic errors $\delta\mu_{J_{\ell}}$ per degree due to the mismodelling $\delta J_2, \delta J_4, \delta J_6,...$ (MGS95J model) in the uncancelled even zonal harmonics in the combination of  the nodes of probe P1 ($a_1=9500$ km, $i_1=89$ deg, $e_1=0.01$) and probe P2 ($a_2=9600$ km, $60$ deg $\leq i_2\leq 89$ deg, $e_2=0.03$) with \rfr{koff} which cancels out entirely the bias due to $\delta R$.}
   \label{figura14}
   \end{figure}

 An approach that could be  fruitfully followed within the framework of the linear combination strategy with two probes P1 and P2, especially in view of likely future improvements
 in the even zonal harmonics of the areopotential, consists in designing a combination which, by construction, entirely cancels out  the bias due to the uncertainty in $R$ \eqi\delta\mu_R\leq \left|\dot\Omega_{.R}^{\rm P2} + k_1\dot\Omega_{.R}^{\rm P1}\right|\left(\rp{\delta R}{R}\right),\eqf being, instead, affected by $\delta J_2, \delta J_4, \delta J_6$. It is analogous to the three-nodes combination of \rfr{combi}, with
 \eqi k_1 = -\rp{\dot\Omega_{.R}^{\rm P2}}{\dot\Omega_{.R}^{\rm P1}}.\lb{koff}\eqf
 Indeed, as can be noted in Figure \ref{figura14}, $\delta\mu_{J_{\ell}}$ would be very good  ($\lesssim 1\%$), according to the present-day MGS95J model, provided that inclinations $i\ll 90$ deg are adopted for the second probe.
 It turns out that the most relevant even zonals would be $J_4$ and $J_6$ with a bias of the order of $\approx 1-0.5\%$.
With \rfr{koff} an inclination range as large as ten degrees would be admissible: it is really a great advantage with respect to the other cases examined so far.

 %\subsection{Summary}
 %Limiting to a two-probe P1-P2 scenario in high-altitude orbits ($a\approx 9500$ km), we examined two possibilities: using the sum of their nodes to %cancel, in principle, the effect of all the even zonals (Section \ref{supl}), and linearly combining the nodes of the two probes to remove only the %bias due to $\ell=2$ (Section \ref{merda}) or the one by $\delta R$ (Section \ref{frutto}). The latter choice is by far better because it allows for %much more loose constraints on the semimajor axis and the inclination ($a_1=9500$ km, $i_1 = 89$ deg; $a_2 = 9600$ km, $60$ deg $\leq i_2\leq$ 89 %deg) and is more effective in reducing the impact of the uncertainty in the even zonals  ($\delta\mu_{J_{\ell}}\lesssim 1\%$) by completely canceling %out the bias due to the radius $R$.
 %
 \section{Conclusions}
 In this paper in examining the possibility of designing a dedicated mission to measure the Lense-Thirring effect in the gravitational field of Mars by analyzing  the secular precessions of the node of one or more spacecrafts we focussed on the main systematic errors of gravitational origin and strategies to reduce them.

 The main source of systematic bias is the mismodelling in some of the parameters entering the multipolar expansion of the  Newtonian part of the areopotential, especially the Mars' equatorial radius $R$ and, to a lesser but non-negligible extent, the even zonal harmonics $J_{\ell},\ell=2,4,6,..$. Since the resulting aliasing node precessions   are proportional to $R^{\ell}a^{-(3/2 + \ell)}\cos i$, a high-altitude ($a\approx 9500$ km), polar ($i=90$ deg) probe would be, in principle, an optimal solution, but the consideration of the unavoidable orbit injection errors showed that such an option is unfeasible because it would require too stringent constraints on the departures from the ideal case $i=90$ deg.
 It could be possible, in principle, to suitably combine the nodes of such a probe with those of the natural satellites of Mars, Phobos and Deimos. However, the Lense-Thirring shifts of such bodies ($1-10$ cm yr$^{-1}$) are orders of magnitude too small with respect to the present accuracy in reconstructing their orbits ($1-10$ km).

 A viable option consists in using two probes P1 and P2 at high-altitudes ($a_1=9500$ km, $a_2=9600$ km) and different inclinations
 ($i_1\approx 90$ deg, $60$ deg $\leq i_2 \leq 89$ deg), and combining their nodes so as to entirely cancel out the bias due to $R$: the resulting bias due to $\delta J_2, \delta J_4, \delta J_6...$  would be $\lesssim 1\%$, according to the present-day MGS95J gravity model, over a  range of values for the inclination as large as ten degrees.

 A major challenge would certainly be reaching a satisfactory accuracy in reconstructing the orbits of such probes whose Lense-Thirring out-of-plane shifts would amount to about $10$ cm yr$^{-1}$; for example, the spacecraft should remain operative for many years, without any failure in the communication with the Earth, and it is likely that one or more landers would be required as well.

%-----------------------------------------

\end{document}